\documentclass{osa-article}
\usepackage[per-mode = symbol]{siunitx}

\journal{osac}
\articletype{Research Article}
\begin{document}
\title{Broadband SiN Asymmetric Directional Coupler for 840 nm Operation}
\author{Stefan Nevlacsil,\authormark{1,*} Moritz Eggeling,\authormark{1} Paul Muellner,\authormark{1} Guenther Koppitsch,\authormark{2} Martin Sagmeister,\authormark{2} Jochen Kraft,\authormark{2} and Rainer Hainberger\authormark{1}}
\address{\authormark{1}AIT Austrian Institute of Technology GmbH, Giefinggasse 4, 1210, Vienna, Austria\\
\authormark{2}ams AG, Tobelbader Stra\ss e 30, 8141 Premst\"atten, Austria
}
\email{\authormark{*}stefan.nevlacsil.fl@ait.ac.at} %% email address is required
%
%%%%%%%%%%%%%%%%%%% abstract %%%%%%%%%%%%%%%%
\begin{abstract}
Silicon nitride based photonic integrated circuits offer a wavelength operation window in the near infrared down to visible light, which makes them attractive for life science applications. However, they exhibit significantly different behavior in comparison to better established silicon on insulator counterparts due to the lower index contrast. One of the most important building blocks in photonic integrated circuits are broadband couplers with a defined coupling ratio. We present silicon nitride broadband asymmetric directional coupler designs with 50/50 and 90/10 splitting ratios with a central wavelength of \SI{840}{\nano\metre} for both TE- and TM-like polarization. We show that silicon nitride broadband asymmetric directional couplers can be designed accurately in a time efficient way by using a general implementation of the coupled mode theory. The accuracy of the coupled mode theory approach is validated with finite difference time domain simulations and confirmed with measurements of four coupler configurations.
\end{abstract}
%
%%%%%%%%%%%%%%%%%%%%%%%%%%  body  %%%%%%%%%%%%%%%%%%%%%%%%%%
\section{Introduction}
During recent years the effort to utilize photonic integrated circuits (PIC) with multiple functional components for life science applications has been continuously increasing\cite{Hoffman.2016,Munoz.2017,Tinguely.2017}. The wavelength region of major interest for these types of applications is in the visible (VIS) and near infrared (NIR) domain below \SI{1}{\micro\metre}. With a lower transmittance limit of \SI{1.1}{\micro\meter} for silicon \cite{Rahim.2017} the widely used silicon on insulator (SOI) technology platform is not suitable for many life science applications. Silicon nitride (SiN) as guiding material on the other hand exhibits high transmittance values down to \SI{0.4}{\micro \meter} and can also be fabricated in already existing CMOS fabs for electronic integrated circuits. For the deposition of the SiN on the cladding material, typically SiO$_2$, either low pressure chemical vapour deposition (LPCVD) or plasma enhanced chemical vapour deposition (PECVD) can be used. While the LPCVD process requires temperatures above >~\SI{700}{\celsius}, the process temperature for PECVD can be kept low enough (<~\SI{400}{\celsius}) to maintain CMOS compatibility \cite{Gorin.2008, Kaloyeros.2017}. This makes a monolithic co-integration of PECVD SiN waveguides with photodiodes on a single silicon chip feasible and more complex wafer bonding can be avoided, which reduces both costs and risks for packaging. Although PECVD SiN exhibits higher waveguide propagation losses than LPCVD SiN, CMOS-compatible PECVD SiN waveguides have been fabricated with propagation losses <~\SI{1}{\deci\bel\per\centi\metre} \cite{Subramanian.2013,RomeroGarcia.2013, Muellner.2015}.\\
Splitting of optical power into two waveguide paths is one of the most fundamental functions desired in PICs. To ensure a versatile range of applications the splitting ratio has to be selectable and constant over a large wavelength bandwidth. Directional couplers (DC) achieve arbitrary splitting ratios, by adjusting the length of their coupling region, consisting of two closely adjacent waveguides. In the case of single mode operation, for an individual waveguide, two modes are supported for the coupled waveguides, a symmetric and antisymmetric mode. The coupling length for a specific splitting behavior is proportional to the difference of propagation constant of the two modes. Since the propagation constant is strongly dependent on the wavelength \cite{Reider.2016}, the splitting ratio varies significantly across larger bandwidths. Various concepts to reduce the wavelength dependence have been successfully demonstrated in other material systems. This includes concepts using adiabatic width variations of adjacent waveguides \cite{Adar.1992}, bent coupled sections \cite{Chen.2017}, shallow edge tapers \cite{Gupta.2017}, and Mach-Zehnder interferometer (MZI) configurations \cite{Jinguji.1990,Lu.2015}, which are a combination of two symmetric DCs connected via a relative phase shift section. We chose the MZI type as a good compromise between footprint, since they are more compact than adiabatic DCs, and simplicity both in terms of design approach and fabrication demands compared to shallow edge tapers and bent coupling sections.
\section{Mach-Zehnder interferometer type directional coupler}
\begin{figure}[ht!]
	\centering\includegraphics{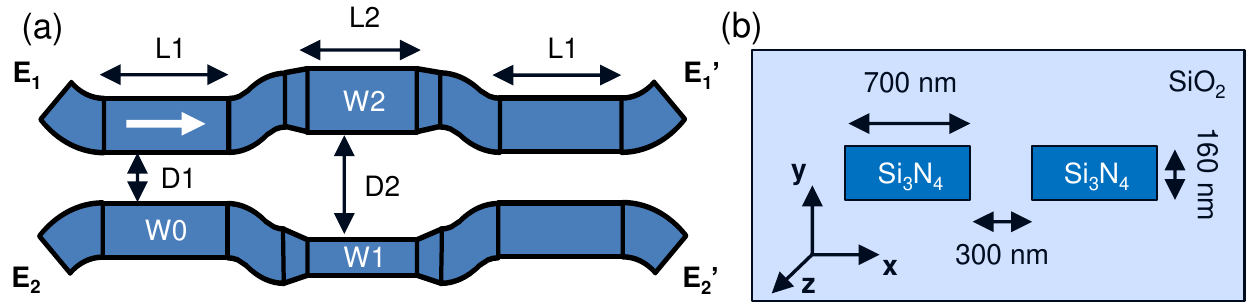}
	\caption{(a) Top view of the chosen design. The waveguide widths were chosen to only support the fundamental mode (W0 = \SI{0.7}{\micro\metre}, W1 = \SI{0.6}{\micro\metre}, W2 = \SI{0.8}{\micro\metre}). At the distance D2 the waveguides are sufficiently decoupled (D1 = \SI{0.3}{\micro\metre}, D2 = \SI{1.5}{\micro\metre}). For the calculations $\text{E}_\text{1}$ is taken as input port with the light propagating from left to right, as indicated by the white arrow. The input, output and s-bends have a radius of \SI{150}{\micro\metre}. The width of the waveguides are changed within a \SI{1}{\micro\metre} long linear taper. (b) Cross section of the symmetric waveguide section of the DC.}
	\label{fig:layout_crosssection}
\end{figure}

There are two simple ways to introduce a relative phase shift in the phase section between the two DCs of an MZI, either by different waveguide lengths or by different propagation constants.
\begin{figure}[ht!]
	\centering\includegraphics{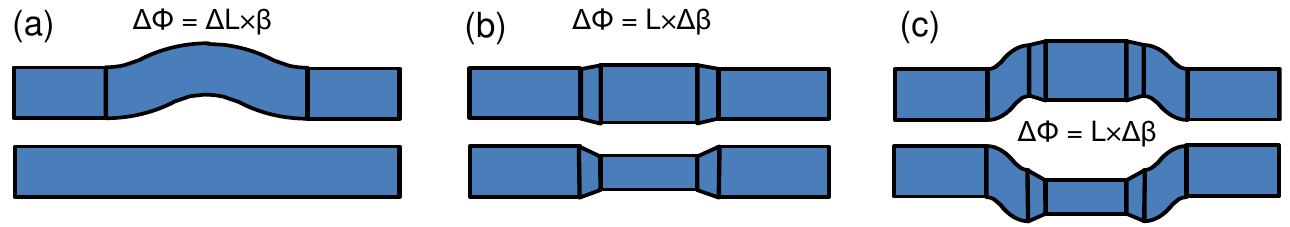}
	\caption{Different implementations of MZI type directional couplers with phase shifts $\Delta\phi$ induced by (a) waveguide length differences or by (b), (c) propagation constant differences.}
	\label{fig:mzi_layouts}
\end{figure}
In the first case (Fig.~\ref{fig:mzi_layouts}a), the phase shift $\Delta\phi$ is proportional to the length difference $\Delta L$ and the propagation constant $\beta$ of the waveguide, while in the latter case (Fig.~\ref{fig:mzi_layouts}b,c) it is proportional to the length $L$ and the difference of the propagation constants $\Delta\beta$ of the two waveguides due to the different widths. Typically, the induced phase shift in the first implementation is much larger than in the second and thus requires a short propagation length difference in the phase section. The downside is that minor fabrication deviations from this ideal length impact the performance significantly. We chose the approach of using waveguides with different widths to avoid this aspect. The design of Fig.~\ref{fig:mzi_layouts}b was employed for the realization of broadband couplers with SOI wire waveguides\cite{Lu.2015}. For this material system the different widths lead to a sufficient decoupling of the DC modes. This results in a relative phase shift section in which no power transfer between the two waveguides occurs. However, for SiN, the waveguide modes do not decouple sufficiently by only adjusting the widths. Therefore, the design of Fig.~\ref{fig:mzi_layouts}c was chosen with an additional separation to achieve the required decoupling before changing the widths. Another aspect which has to be considered with more care for SiN compared to SOI is the larger required bend radius which produces significant coupling outside of the straight sections. The fabricated design is shown in Fig.~\ref{fig:layout_crosssection}. \\
In our work, we fixed the waveguide core thickness to $\SI{160}{\nano\meter}$ to achieve low propagation losses while keeping the minimum bend radii small enough to allow the integration of several functional waveguide building blocks on a single PIC. Since systematic studies showed that scattering of light in the PECVD SiN waveguide core is the main source of propagation loss, the waveguide core thickness should be made small. On the other hand, this results in a lower confinement of the guided mode in the waveguide core, which necessitates larger bending radii. We confirmed theoretically and experimentally that for the given waveguide cross section of $\SI{700}{\nano\meter}\times\SI{160}{\nano\meter}$ a bend radius as narrow as $\SI{150}{\micro\meter}$ can be used for both TE- and TM-like polarization without inducing  additional bending losses.  Furthermore, to facilitate efficient end facet coupling to an optical fiber, inverted tapers with square shaped cross sections are used to match the mode fields of the fiber and the waveguide. Also with this respect, the waveguide core thickness of $\SI{160}{\nano\meter}$ provides a good compromise.\\
The coupling gap of $\SI{300}{\nano\meter}$ is set to the minimum distance within a safety margin to the allowed distance between waveguides due to fabrication limits to reduce the device length. Finally, the maximum and minimum widths were set to $\SI{800}{\nano\meter}$ and $\SI{600}{\nano\meter}$, respectively. The width disparity of $\SI{200}{nm}$ allows a reasonable phase shift $\Delta\phi$, while avoiding long taper sections, larger separation (>$\SI{1.5}{\micro\metre}$) in the phase shift section and multimode operation in the wide waveguide section  (>$\SI{800}{\nano\meter}$).
\section{Coupled mode theory}
We introduce a design approach that avoids the use of computationally intensive 3D finite difference time domain (FDTD) simulations. It describes the complete range of fully coupled, decoupled and intermediate states of the device sections. The approach is based on the general form of the coupled mode theory (CMT)\cite{Reider.2016} and the simplified form used for SOI waveguides \cite{Lu.2015}. The CMT describes the interaction of two modes propagating in straight sections of two adjacent waveguides with a coupling matrix. In the limit of weak coupling it is defined as
\begin{equation}
C = \begin{pmatrix}
t & -i k \\
-i k & t^*
\end{pmatrix} \text{exp}(i\phi)
\label{eq:coupling_matrix}
\end{equation}	
with the parameters
\begin{equation}
t = \text{cos}(Kz) - i \frac{\Delta\beta}{K} \text{sin}(K z), \\
\hspace{0.5cm} k = \frac{\kappa}{K}\text{sin}(K z), \\
\hspace{0.5cm} \phi = \bar{\beta} z,
\label{eq:matrix_element}
\end{equation}
the mean and difference of  $\beta_1 \text{and } \beta_2$ of the individual uncoupled waveguide modes
\begin{equation}
\bar{\beta} = \frac{\beta_1+\beta_2}{2}, \hspace{0.5cm} \Delta\beta = \frac{\beta_1-\beta_2}{2},
\label{eq:beta_uncoupled}
\end{equation}
and the interaction section length $z$, the   coupling $K$ and coupling coefficient $\kappa$. The propagation constant is defined as
\begin{equation}
	\beta(\lambda) = \frac{2 \pi}{\lambda} n_{\textnormal{eff}}(\lambda).
	\label{eq:beta_neff}
\end{equation}
The propagation constant of the modes in the coupled waveguides is defined as
\begin{equation}
	\beta^\pm = \bar{\beta} \pm K \\ 
	\label{eq:beta_coupled_single}
\end{equation}
with the additional relation that
\begin{equation}
	K = [(\Delta\beta)^2+|\kappa|^2]^{1/2}.
	\label{eq:effective_coupling}
\end{equation}
With the difference of the coupled mode propagation constants
\begin{equation}
\Delta\beta_{\textnormal{coup}} = \frac{\beta^+ -\beta^-}{2},
\label{eq:beta_coupled}
\end{equation}
Eq.~\eqref{eq:beta_coupled_single} can be rewritten to show that $K = \Delta\beta_{\textnormal{coup}}$. Rewriting Eq.~\eqref{eq:effective_coupling} leads to the relation
\begin{equation}
	\kappa = [K^2-(\Delta\beta)^2]^{1/2} = [(\Delta\beta_{\textnormal{coup}})^2-(\Delta\beta)^2]^{1/2}.
	\label{eq:kappa}
\end{equation}
Therefore, to calculate the coupling matrix for a section length $z$ only the propagation constants of both the uncoupled ($\beta_1 \text{, } \beta_2$) and coupled modes ($\beta_+ \text{, } \beta_-$) for a given cross section are required. This can be done for any arbitrary width arrangement of the waveguide pair to describe the DC completely. 
\section{Design and results}
\label{sec:Design and results}
\begin{figure}[ht!]
	\centering\includegraphics{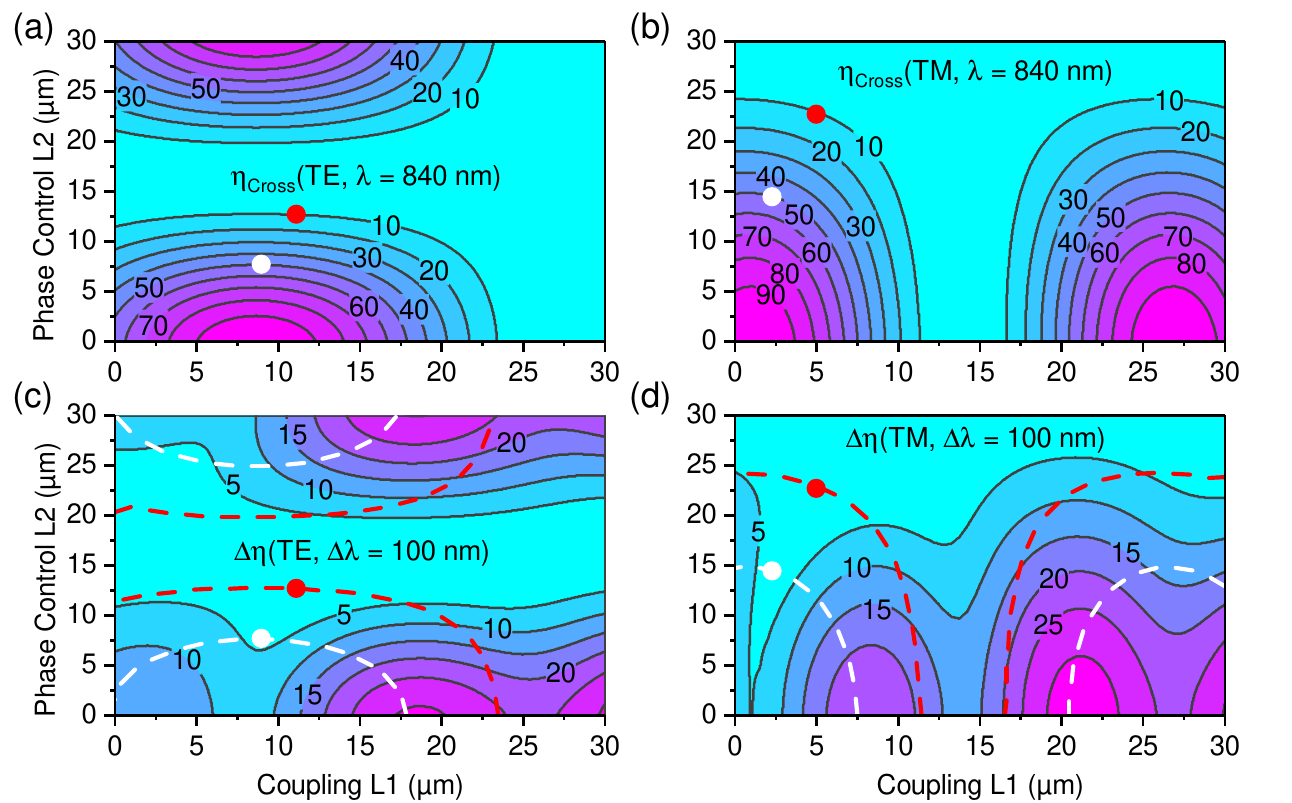}
	\caption{Splitting ratio in percent of the optical power at the cross port ($\text{E}_\text{2}'$) for a parameter sweep of L1 and L2 from \SIrange{0}{30}{\micro\metre} for (a) TE- and (b) TM-like polarization. The red and white dots indicate the device parameters chosen for 3D FDTD simulation and fabrication. Optimization parameter $\Delta\eta$ in percent of the splitting ratio across the bandwidth of \SI{100}{\nano\metre} for (c) TE- and (d) TM-like polarization. The red and white dashed lines indicate the 50/50 and 90/10 splitting ratio, respectively.}
	\label{fig:sbend_width_0nm}
\end{figure}

\begin{figure}[ht!]
	\centering\includegraphics{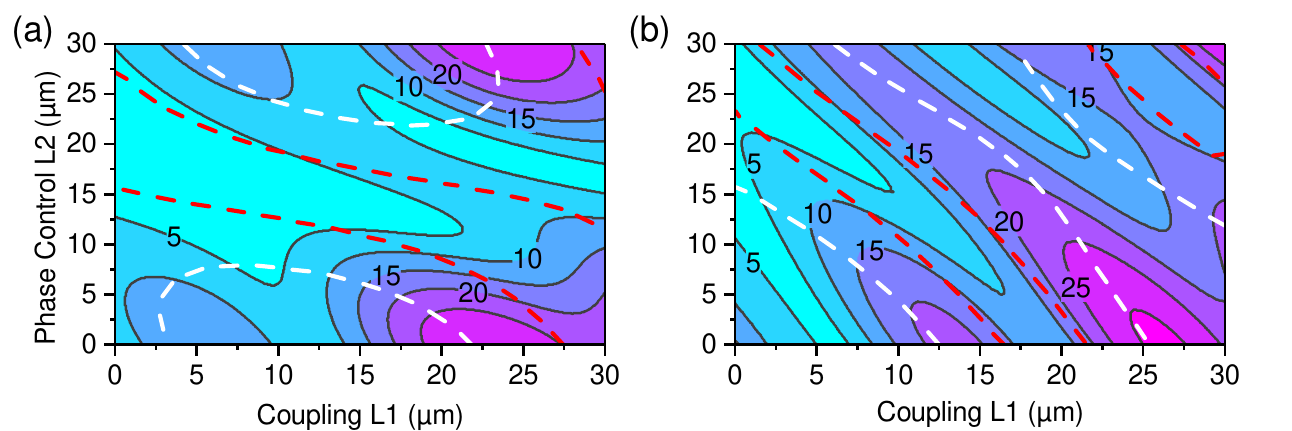}
	\caption{Optimization parameter $\Delta\eta$ in percent for a DC without added separation for (a) TE- and (b) TM-like polarization. The red and white dashed lines indicate the 50/50 and 90/10 splitting ratio, respectively. For TM-like polarization the mixture between coupling and phase shifts results in a significant distortion in the $\Delta\eta$ over the wavelength bandwidth. Therefore, no adequate splitting behavior was found and a separation was added.}
	\label{fig:taper_width_0nm_TM_var}
\end{figure}

\begin{figure}[ht!]
	\centering\includegraphics{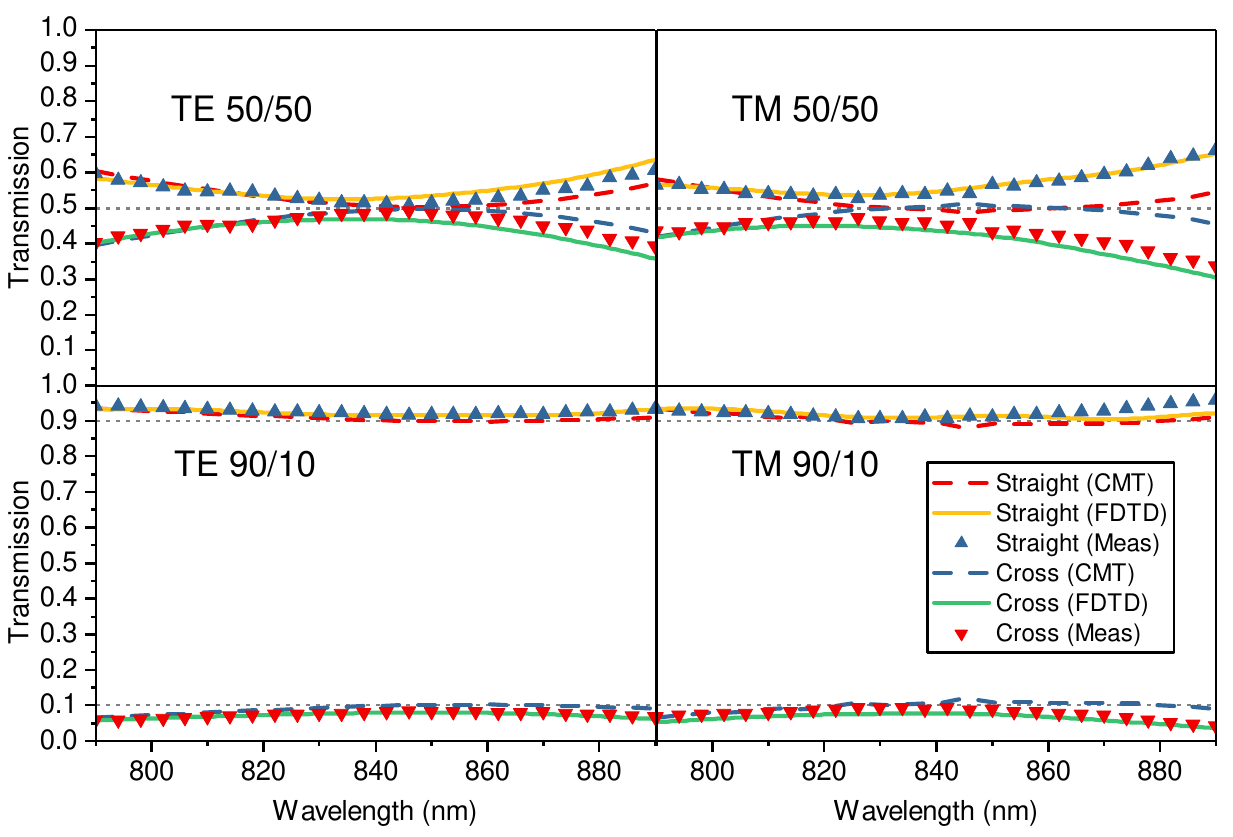}
	\caption{Splitting ratio behavior for 50/50 and 90/10 MZI DCs for TE- and TM-like polarization. The gray dashed line indicates the intended splitting ratio. The measured light power at the output ports was normalized to the total output power of both ports. 3D FDTD simulations and the measurements show almost identical behavior with small deviations which can be attributed to fabrication limits.}
	\label{fig:splitting_ratio}
\end{figure}

\begin{figure}[ht!]
	\centering\includegraphics{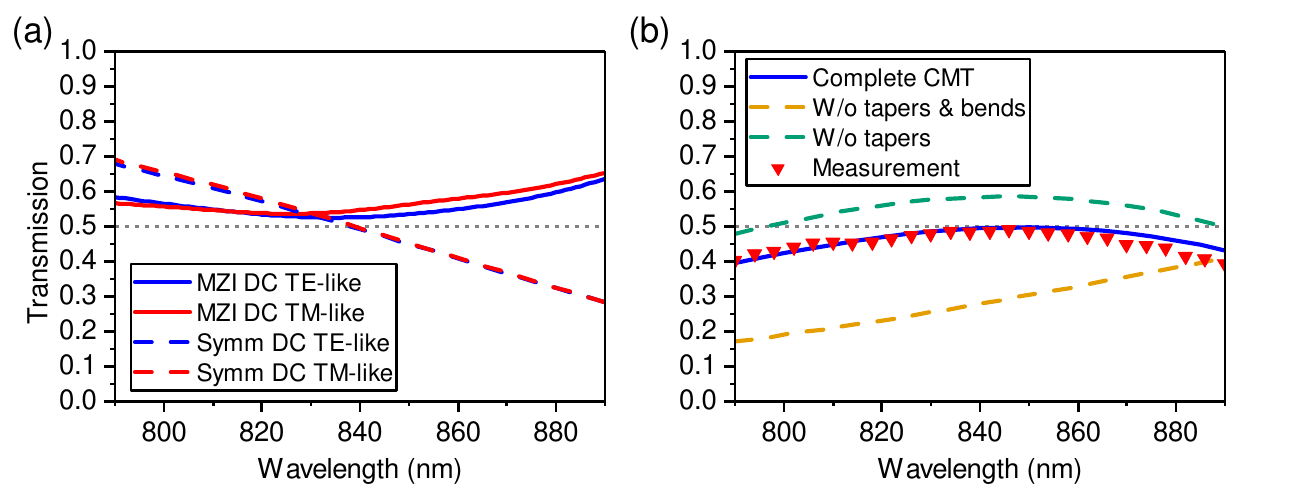}
	\caption{(a) 3D FDTD simulations of the power transfer ratio to the straight port for a 50/50 DC with a standard symmetric design and the broadband MZI design for TE and TM-like polarization. The MZI DC design significantly improves the broadband behavior compared to the standard symmetric design over a bandwidth of $\SI{100}{\nano\metre}$. In the standard symmetric DC the distance between the coupling waveguides was set to $\SI{300}{\nano\metre}$ for the TE-like polarization, while for the TM-like polarization the distance had to be increased to $\SI{500}{\nano\metre}$ to reduce the coupling in the input and output bent sections. (b) Comparison of CMT simulation results and the measurement for the 50/50 MZI DC for TE-like polarization for the cross port. This shows that by taking into account the coupling effect in the bent and tapered sections (complete CMT), the CMT closely resembles the measurement, while omitting this effect either in the tapered sections (w/o tapers) or in both the tapered and bent sections (w/o tapers \& bends) results in significant deviations.}
	\label{fig:conventional_coupler}
\end{figure}

To describe the coupling behavior throughout a device with multiple distinct cross sections, the coupling matrices are multiplied in sequence to an input vector $(E_1,E_2)^T$. For the calculation of $n_{\textnormal{eff}}$ the eigenmode solver "Mode Solutions" \cite{LumericalInc..2018} was used. The bends and tapers were divided into 21 equally long subsections. Coupling effects in the input and output bent sections were considered up to a separation of $\SI{3}{\micro\meter}$, which is well above the coupling threshold. For each distinct cross section 21 wavelengths with a central wavelength of $\lambda = \SI{840}{\nano\metre}$ and a bandwidth of \SIrange{790}{890}{\nano\metre} were simulated. The material refractive indices were determined with ellipsometry measurements. The normalized power at the straight and cross output port for an input vector of $(E_1,E_2)^T = (1,0)$ are calculated with
\begin{equation}
	\eta_{\textnormal{cross}} = \frac{|E_2'|^2}{|E_1'|^2+|E_2'|^2} \hspace{2cm} \eta_{\textnormal{straight}} = \frac{|E_1'|^2}{|E_1'|^2+|E_2'|^2}.
	\label{eq:eta}
\end{equation}
The section lengths L1 and L2 are chosen to achieve the lowest mean squared error ($\Delta\eta$) for a specific splitting ratio. The $\Delta\eta$ is calculated with
\begin{equation}
\Delta\eta = \frac{1}{21}\sum_{n=1}^{21} [\eta(\lambda_n) - \eta(\lambda = \SI{840}{\nano \metre})]^2.
\label{eq:eta_variance}
\end{equation}	
Figure~\ref{fig:sbend_width_0nm} shows the contour plots for the section lengths parameter sweep.
To show the quality of the CMT design approach and the capability of the MZI DC the splitting ratios 50/50 and 90/10 for TE- and TM-like polarization were chosen for fabrication. Splitting ratios other than 50/50 are expected to exhibit better broadband behavior, with the possibility of switching the input port for higher amounts of light required at the cross port. Figure~\ref{fig:taper_width_0nm_TM_var} depicts the $\Delta\eta$ for the MZI DC without an additional separation of the waveguides in the phase shift section. Especially for TM-like polarization the coupling within the phase shift section induces a strong degradation of the broadband behavior. This clearly shows that an additional separation is necessary.
The PICs comprising the chosen devices were fabricated by ams AG using their production line \cite{Muellner.2015}. To maintain CMOS compatibility the SiN thin film was deposited using PECVD. The structural components were fabricated with deep ultraviolet photolithography and reactive ion etching. The photonic devices on chip were evaluated via fiber end face coupling with a tunable Ti-sapphire laser. The power was directly applied to the input ports individually and measured at both output ports. With all port combinations a mean value was calculated for the splitting ratios of each device. Figure~\ref{fig:splitting_ratio} summarizes the simulation results of CMT and 3D FDTD \cite{LumericalInc..2018} compared to the measurement of the fabricated device.
Figure~\ref{fig:conventional_coupler}a compares 3D FDTD simulations of the 50/50 broadband MZI DC designs and  standard symmetric designs with a waveguide width of $\SI{700}{\nano\metre}$ for TE and TM-like polarization. In the standard symmetric DC the distance between the coupling waveguides was set to $\SI{300}{\nano\metre}$ for the TE-like polarization, while for the TM-like polarization the distance had to be increased to $\SI{500}{nm}$ to reduce the coupling in the input and output bent sections. For the through port of the broadband MZI DC the power stays within a window of approximately 10\% in comparison to a power variation of 40\% for the standard DC design. \\Figure~\ref{fig:conventional_coupler}b illustrates the importance of taking the coupling in the bent and tapered sections into account by comparing the complete and partial CMT simulations with the actual measurement. The complete CMT simulation, which includes the coupling effect in the bent and tapered sections, closely resembles the measurement. On the other hand, omitting the coupling effect in both the bent and the tapered sections diminishes the broadband behavior and shifts the splitting ratio, while omitting the coupling in the tapered sections only produces a constant offset of the ratio. This leads to the conclusion that by changing the length L2 the coupling ratio can be fine tuned without extensive degradation of the broadband behavior. In this work, the main goal was to closely achieve a 50/50 ratio for the central wavelength of $\SI{840}{nm}$.
\section{Conclusion}
The PECVD-SiN waveguide based MZI DCs exhibit good broadband performance for the chosen splitting ratios of 50/50 and 90/10 in the wavelength range of \SIrange{790}{890}{\nano\metre} for TE- and TM-like polarization. Other splitting ratios can be obtained using the presented design maps. The implemented CMT formalism can be employed to calculate the splitting behavior of the MZI DC with low computational requirements in comparison to 3D FDTD simulations. For the simulation of a single device CMT requires approximately a fifth of the process time of 3D FDTD. Each additional length variation can be done with CMT in seconds compared to typically multiple hours for a single 3D FDTD simulation of another device. The good agreement with experimental results confirms that the CMT approach is a viable option for the device design. To our knowledge this is the first demonstration of SiN waveguide based broadband couplers with selectable splitting ratio in the <~\SI{1}{\micro\metre} NIR wavelength region. These couplers open the way for PIC applications in this wavelength regime that benefit from selectable power distribution across a broad wavelength range, e.g. optical coherence tomography where more power is required at the low reflective biological sample than in the reference path.
\section*{Funding}
This research has received funding from the European Union's Horizon 2020 research and innovation program under grant agreement No 688173 (OCTCHIP).
\section*{Acknowledgments}
Portions of this work were presented at the European Conference on Integrated Optics in 2018, ''Broadband SiN Asymmetric Directional Coupler for 850 nm Wavelength Region''. The ellipsometry measurements were provided by Stefan Partel at the FH Vorarlberg University of Applied Sciences.
%%%%%%%%%%%%%%%%%%%%%%% References %%%%%%%%%%%%%%%%%%%%%%%%%
%%%%%%%%%% If using BibTeX:
\bibliography{refs}
\end{document}